\documentclass[12pt,a4paper]{article}

\usepackage{epsfig}
\usepackage{amsmath,amsfonts,amssymb}
\usepackage{cite}
\usepackage{colortbl}
\usepackage{pifont}

\providecommand{\openone}{\leavevmode\hbox{\small1\kern-3.8pt\normalsize1}}

\newcommand{\vl}{V_L}
\newcommand{\vr}{V_R}
\newcommand{\gl}{g_L}
\newcommand{\gr}{g_R}

\newcommand{\RE}{\text{Re}}

\begin{document}

\begin{center}
\begin{Large}
{\bf Model-independent measurement \\[1mm] of the top quark polarisation}
\end{Large}

\vspace{0.5cm}
J. A. Aguilar--Saavedra$^{a,b}$, R. V. Herrero-Hahn$^c$ \\[1mm]
\begin{small}
{\it $^a$ Departamento de F\'{\i}sica Te\'orica y del Cosmos, 
Universidad de Granada, Granada, Spain} \\
{\it $^b$ Instituto de F\'{\i}sica de Cantabria (CSIC-UC), Santander, Spain} \\
{\it $^c$ Departamento de F\'{\i}sica Aplicada, Universidad de Granada, Granada, Spain}
\end{small}
\end{center}

\begin{abstract}
We introduce a new asymmetry in the decay $t \to Wb \to \ell \nu b$, which is shown to be directly proportional to the polarisation of the top quark along a chosen axis, times a sum of $W$ helicity fractions. The latter have already been precisely measured at the Tevatron and the Large Hadron Collider. Therefore, this new asymmetry can be used to obtain a model-independent measurement of the polarisation of top quarks produced in any process at hadron or lepton colliders.
\end{abstract}

\section{Introduction}

Precision measurements of the top quark properties offer an excellent opportunity to explore indirect effects of new physics beyond the Standard Model (SM). Their theoretical interest is motivated by the large top quark mass, which leads to the common belief that this fermion may be quite sensitive to new physics effects. And, on the experimental side, top quark studies are greatly facilitated by the short lifetime of this quark, $\tau \sim 4 \times 10^{-25}$~s, which prevents complications from hadronisation effects and allows to study in detail the properties of a ``bare'' quark. Thus, for example, the $W$ helicity fractions~\cite{Kane:1991bg} have been precisely measured at the Tevatron and the Large Hadron Collider (LHC)~\cite{Abazov:2010jn,Aaltonen:2012tk,Aad:2012ky,CMS},
namely the relative fractions of $W$ bosons with helicity $\pm 1,0$ produced in the decay $t \to Wb$.

New physics can enter both the production and decay of the top quark. New production mechanisms may be difficult to spot directly, as is the case of wide $t \bar t$ resonances~\cite{Barcelo:2011vk}, superpartners~\cite{Hikasa:1999wy} or non-resonant~\cite{Degrande:2010kt} contributions, the latter including $t$-channel flavour-changing processes~\cite{Jung:2009jz,Han:1998tp}. But the presence of such contributions (arising for example in models addressing the anomalous Tevatron $t \bar t$ asymmetry) would generally result in a top polarisation or $t \bar t$ spin correlation~\cite{Cao:2010nw} different from the SM predictions~\cite{Mahlon:1996pn,Mahlon:1999gz,Mahlon:1995zn,Bernreuther:2004jv,Mahlon:2010gw,Bernreuther:2010ny}. These are not measurable quantities, however, and can only be probed by analysing angular distributions in the top decay $t \to Wb \to \ell \nu b$, with $\ell = e,\mu$. (Hadronic $W$ decays and leptonic decays to taus are sensitive to the top polarisation too, but their experimental measurement is much more difficult.) For example, a well-known method to probe the top polarisation is through the angular distributions of its decay products in the top quark rest frame. These take the form
\begin{equation}
\frac{1}{\Gamma} \frac{d\Gamma}{d\cos\theta_X} = \frac{1}{2} ( 1 + P_z \alpha_X \cos \theta_X ) \,,
\label{ec:dist1}
\end{equation}
being $\theta_X$ the angle between the momentum $\vec p_X$ of the decay product $X=\ell,\nu,b,W$ in the top quark rest frame, and an arbitrary direction $\hat z$ chosen to quantise the top spin. In the above equation, $P_z$ is the top polarisation along this direction and $\alpha_X$ are constants called ``spin analysing power'' of the particle $X$, which can be affected by top anomalous couplings~\cite{Jezabek:1994zv,AguilarSaavedra:2006fy}. (Radiative corrections to these quantities have been computed in~\cite{Brandenburg:2002xr,Bernreuther:2004jv,Groote:2006kq}.) Hence,
Eq.~(\ref{ec:dist1}) clearly shows a production-decay interplay in the $\theta_X$ distributions: the measurable quantities are the products $P_z \alpha_X$, which depend on the production ($P_z$) and decay properties ($\alpha_X$) of the top quark. This is a general feature: since the top polarisation (as well as the $t \bar t$ spin correlation) can only be measured through top decay distributions, the resulting observables are also sensitive to anomalous contributions to the  $Wtb$ vertex.\footnote{We ignore here other types of new physics in the top decay, such as the rare modes, which give rise to different final states, often easily identifiable, and assume that the interaction between the $W$ boson and the charged leptons is the SM one, as implied by low energy measurements.} A non-trivial but important issue is then to disentangle new physics in production and decay. This, of course, would become crucial in case that a deviation from the SM predictions was found.

Previous literature~\cite{Grzadkowski:2002gt,Grzadkowski:2001tq,Rindani:2000jg,Godbole:2006tq,Godbole:2010kr,Godbole:2011vw} has attempted to get rid of the dependence on the top decay vertex by noting that for the charged lepton the spin analysing power $\alpha_\ell$ depends on $Wtb$ anomalous couplings only quadratically, so $\alpha_\ell$ {\it should be} less sensitive to new physics. This solution is not satisfactory, however, not only because the new physics affecting top production may modify $P$ at quadratic level too,\footnote{This is always the case for non-interfering new physics, for example involving flavour-changing neutral currents or charged-current interactions with light quarks~\cite{AguilarSaavedra:2010sq}.} but also because anomalous $Wtb$ couplings are not sufficiently constrained from other sources so as to imply that their quadratic contributions are small. In~\cite{AguilarSaavedra:2010nx} it has been shown that a global fit to several top decay observables (including $W$ helicity fractions, $\alpha_\ell$ and $\alpha_b$) can be used to extract $P_z$ from single top and $t \bar t$ measurements. In this Letter we focus on a more direct measurement of the top polarisation and introduce a ``doubly forward-backward'' top decay asymmetry
\begin{equation}
A_{FB}^{tW} = \frac{N(\cos \theta \times \cos \theta^* > 0) - N(\cos \theta \times \cos \theta^* < 0)}{N(\cos \theta \times \cos \theta^* > 0) + N(\cos \theta \times \cos \theta^* < 0)}  \,,
\label{ec:Adef}
\end{equation}
where $N$ stands for the number of events; $\theta$ is the angle between the $W$ momentum in the top rest frame $\vec p_W$ and a chosen top spin quantisation axis $\hat z$; $\theta^*$ is the angle between the charged lepton momentum in the $W$ rest frame, $\vec{p_\ell^*}$, and $\vec p_W$. We show that this asymmetry is related to the top polarisation along the $\hat z$ direction and the $W$ helicity fractions $F_i$ by
\begin{equation}
A_\text{FB}^{tW} = \frac{3}{8} P_z \left( F_+ + F_- \right)
\label{ec:rel}
\end{equation}
in full generality. Since the $W$ helicity fractions can be (and have actually been) measured in a model-independent fashion, $A_\text{FB}^{tW}$ provides a model-independent measurement of the top polarisation along a chosen axis, in any process of top production at hadron or lepton colliders. In addition, we present here an inequality involving $\alpha_W$ and $W$ helicity fractions, which can be used to obtain lower bounds on $P_z$ from the measurement of the $\cos \theta_W$ distribution.

\section{Top quark decay in the helicity formalism}

We use the Jacob-Wick helicity formalism~\cite{Jacob:1959at} (see also~\cite{Chung:1971ri}) to describe the decay of the top quark and $W$ boson using general arguments of angular momentum conservation. Let us fix a $(x,y,z)$ coordinate system in the top quark rest frame, with the positive $z$ axis along the direction in which we want to quantise the top spin.
The most general spin state of an ensemble of top quarks can be described by a density matrix
\begin{equation}
\rho = \frac{1}{2} \left( \! \begin{array}{cc} 1+P_z & (P_x + i P_y)  \\ (P_x - i P_y) & 1-P_z \end{array} \! \right) \,,
\label{ec:rho}
\end{equation}
with $P_i = 2 \langle S_i \rangle$. We do not specify the orientation of our $x$ and $y$ axes, which is not relevant for our discussion. The amplitudes for the decay $t \to Wb$, for a top quark having third spin component $M=\pm 1/2$ and the $W$ boson and $b$ quark having helicities $\lambda_1=\pm 1,0$, $\lambda_2 = \pm 1/2$, respectively, can be written as
\begin{equation}
A_{M \lambda_1 \lambda_2} = a_{\lambda_1 \lambda_2} D_{M \Lambda}^{1/2\,*}(\phi,\theta,0) \,,
\end{equation}
being $(\theta,\phi)$ the polar and azimuthal angles of $\vec p_W$ in the $(x,y,z)$ coordinate system, $\Lambda=\lambda_1-\lambda_2$ and
\begin{equation}
D^j_{m'm}(\alpha,\beta,\gamma) \equiv \langle jm' | e^{-i \alpha J_z} e^{-i \beta J_y} e^{-i \gamma J_z} | jm \rangle
\end{equation}
a Wigner function for a rotation $R(\alpha,\beta,\gamma)$ parameterised by its Euler angles (explicit expressions for low $j$ can be found in~\cite{Beringer:1900zz}). Hence, we see that all the dependence on $M$ and the direction of $\vec p_W$ is encoded in the $D_{M\Lambda}^{1/2}$ function, while $a_{\lambda_1 \lambda_2}$ only depend on the helicities, invariant under rotations. There are only eight non-zero amplitudes, corresponding to $M=\pm 1/2$ and
\begin{equation}
a_{-1\, -1/2} \,,~ a_{0\, -1/2} \,,~ a_{0\, 1/2} \,,~ a_{1\, 1/2} \,,
\end{equation}
because the two remaining helicity combinations imply a total angular momentum $\pm 3/2$ of the $Wb$ pair  along the direction of $\vec p_W$, which is forbidden for a spin-$1/2$ decaying top quark.

The decay $W \to \ell \nu$ can also be described in a similar fashion, introducing a $(x',y',z')$ coordinate system in the $W$ boson rest frame, with the $z'$ axis in the direction of $\vec p_W$. Then,
the full decay amplitude can be written as
\begin{equation}
A_{M \lambda_2 \lambda_3 \lambda_4} = \sum_{\lambda_1} a_{\lambda_1 \lambda_2} b_{\lambda_3 \lambda_4} D_{M \Lambda}^{1/2\,*}(\phi,\theta,0) \, D_{\lambda_1 \lambda}^{1\,*}(\phi^*,\theta^*,0) \,,
\label{ec:A}
\end{equation}
with $\lambda_3$ ($\lambda_4$) the helicity of the charged lepton (neutrino) and $\lambda = \lambda_3 - \lambda_4$; $(\theta^*,\phi^*)$ are the polar and azimuthal angles of the charged lepton momentum $\vec{p_\ell^*}$ in the $W$ boson rest frame, using the $(x',y',z')$ coordinate system. (We denote quantities in the $W$ boson rest frame with asterisks, as opposed to quantities in the top quark rest frame.) Notice the coherent sum over $W$ boson helicities $\lambda_1$. In the case of a $W^+$ boson decay, the left-handed structure of the vertex implies $\lambda_3 = 1/2$ for the positively charged lepton (anti-fermion) and $\lambda_4=-1/2$ for the neutrino, both taken massless. 

From Eqs.~(\ref{ec:rho}) and (\ref{ec:A}), the fully differential decay width is
\begin{eqnarray}
\frac{d\Gamma}{d\phi\,d\!\cos\theta\,d\phi^*\,d\!\cos\theta^*} & = & C \sum_{M M' \lambda_1 \lambda_1' \lambda_2} \rho_{MM'} a_{\lambda_1 \lambda_2} a_{\lambda_1' \lambda_2}^* |b_{\lambda_3 \lambda_4}|^2 D_{M\Lambda}^{1/2*} (\phi,\theta,0) D_{M'\Lambda'}^{1/2} (\phi,\theta,0)  \notag \\
& & \times D_{\lambda_1 \lambda}^{1*} (\phi^*,\theta^*,0)  D_{\lambda_1' \lambda}^{1} (\phi^*,\theta^*,0) \,,
\end{eqnarray}
with $C$ a constant phase-space factor and $\Lambda' = \lambda_1'-\lambda_2$. Integrating over the azimuthal angles $\phi$, $\phi^*$ gives factors $2\pi \delta_{MM'}$ and $2\pi \delta_{\lambda_1 \lambda_1'}$, respectively, so that the differential width in the two polar angles reads  
\begin{equation}
\frac{d\Gamma}{d\!\cos\theta\,d\!\cos\theta^*} = 4 \pi^2 C |b_{\lambda_3 \lambda_4}|^2 \sum_{M \lambda_1 \lambda_2} \rho_{MM} |a_{\lambda_1 \lambda_2}|^2
 \left[ d_{M\Lambda}^{1/2} (\theta) \, d_{\lambda_1 \lambda}^{1} (\theta^*) \right]^2 \,,
\label{ec:dG2}
\end{equation}
with
\begin{equation}
d_{m'm}^j(\beta) \equiv \langle jm' | e^{-i \beta J_y} | jm \rangle \,.
\end{equation}
(This distribution has already been obtained explicitly~\cite{Fischer:1998gsa,Fischer:2001gp} within the SM, including radiative corrections.)
The total width for $t \to Wb$ is obtained by integration over the remaining angles,
\begin{equation}
\Gamma = \frac{8 \pi^2}{3} C |b_{\lambda_3 \lambda_4}|^2 \left\{
|a_{-1\, -1/2}|^2 + | a_{0\, -1/2}|^2  + |a_{0\, 1/2}|^2 + |a_{1\, 1/2}|^2
\right\} \,.
\label{ec:G}
\end{equation}
We can identify the helicity fractions $F_{\pm,0}$, as the relative widths for $t \to Wb$ with $\lambda_1 = \pm 1,0$, respectively. Denoting for brevity the sum between brackets in Eq.~(\ref{ec:G}) as $D$, we have
\begin{eqnarray}
F_+ & = & |a_{1\, 1/2}|^2 / D \,, \notag \\
F_0 & = & \left[ | a_{0\, -1/2}|^2  + |a_{0\, 1/2}|^2 \right] / D \,, \notag \\[1mm]
F_- & = & |a_{-1\, -1/2}|^2 / D \,.
\label{ec:F}
\end{eqnarray}
Then, integrating Eq.~(\ref{ec:dG2}) in the four quadrants $\cos \theta \gtrless 0$, $\cos \theta^* \gtrless 0$ and dividing by $\Gamma$ we obtain an explicit expression for the asymmetry in Eq.~(\ref{ec:Adef}),
\begin{equation}
A_\text{FB}^{tW} = \frac{3}{8} P_z \left[ |a_{-1\, -1/2}|^2 + |a_{1\, 1/2}|^2 \right] / D = \frac{3}{8} P_z \left[ F_+ + F_- \right] \,.
\label{ec:rel2}
\end{equation}
For anti-top decays $\lambda_3 = -1/2$ for the negatively charged lepton and $\lambda_4=1/2$ for the neutrino, so $\lambda=-1$ and the resulting asymmetry is
\begin{equation}
\bar A_\text{FB}^{tW} = -\frac{3}{8} P_z \left[ |\bar a_{-1\, -1/2}|^2 + |\bar a_{1\, 1/2}|^2 \right] / D = - \frac{3}{8} P_z \left[ \bar F_+ + \bar F_- \right] \,,
\end{equation}
with the helicity fractions for anti-top decays (denoted with bars, as the corresponding anti-top decay amplitudes) satisfying $\bar F_0 = F_0$, $\bar F_\pm = F_\mp$~\cite{AguilarSaavedra:2010nx}.

A by-product of our analysis is obtained by integrating Eq.~(\ref{ec:dG2}) over the full $\theta^*$ range to obtain the $W$ boson angular distribution, namely Eq.~(\ref{ec:dist1}) for $X=W$. The spin analysing power of the $W$ boson is found to be
\begin{equation}
\alpha_W = \left[ |a_{1\, 1/2}|^2 + | a_{0\, -1/2}|^2  - |a_{0\, 1/2}|^2  -|a_{-1\, -1/2}|^2  \right] / D \,.
\end{equation}
This implies, given Eqs.~(\ref{ec:F}) for the helicity fractions, that
\begin{equation}
\alpha_W \leq F_0 - F_- + F_+ \,.
\label{ec:ineq}
\end{equation}
This inequality is practically saturated in the SM because amplitudes with $\lambda_2=1/2$ are suppressed due to the left-handed $W t_L b_L$ interaction. 

Finally, by integrating Eq.~(\ref{ec:dG2}) over $\theta$ we obtain the well-known $\theta^*$ distribution
\begin{equation}
\frac{1}{\Gamma} \frac{d \Gamma}{d\!\cos \theta^*} = \frac{3}{8}
(1 + \cos \theta^*)^2 \, F_+ + \frac{3}{8} (1-\cos \theta^*)^2 \, F_-
+ \frac{3}{4} \sin^2 \theta^* \, F_0 \,.
\label{ec:dist2}
\end{equation}
In particular, the forward-backward (FB) asymmetry in the $W$ rest frame~\cite{Lampe:1995xb,delAguila:2002nf} is
\begin{equation}
A_\text{FB} = \frac{3}{4} \left[ F_+ - F_- \right] \,.
\label{ec:afb}
\end{equation}
These results help clarify the relation~(\ref{ec:rel}): the subtraction of events with $\cos \theta^* > 0$ and $\cos \theta^* < 0$ removes the contribution from the ``symmetric'' amplitudes $|a_{0\, -1/2}|^2$, $|a_{0\, 1/2}|^2$ entering $F_0$, while the subtraction of events with $\cos \theta > 0$ and $\cos \theta < 0$ removes the polarisation-independent term. 

\section{Discussion}

We have introduced a FB asymmetry using two angles $\theta$, $\theta^*$ in the top quark and $W$ boson rest frames, respectively, showing that it is related to the $W$ helicity fractions by Eq.~(\ref{ec:rel}). Its value in the SM can be computed using previous calculations for the helicity fractions~\cite{AguilarSaavedra:2010nx,Drobnak:2010ej}. The most general effective $Wtb$ interaction arising from dimension-six operators can be parameterized as~\cite{AguilarSaavedra:2008zc}
\begin{eqnarray}
\mathcal{L}_{Wtb} & = & - \frac{g}{\sqrt 2} \bar b \, \gamma^{\mu} \left( \vl
P_L + \vr P_R
\right) t\; W_\mu^- \nonumber \\
& & - \frac{g}{\sqrt 2} \bar b \, \frac{i \sigma^{\mu \nu} q_\nu}{M_W}
\left( \gl P_L + \gr P_R \right) t\; W_\mu^- + \mathrm{h.c.} \,,
\label{ec:lagr}
\end{eqnarray}
being $V_L = V_{tb}$ and $V_R = g_L = g_R = 0$ in the SM. For this general vertex, we obtain
\begin{equation}
A_\text{FB}^{tW} = \frac{3}{4} P_z \frac{B_0}{A_0 + 2 B_0} \,,
\end{equation}
with
\begin{align}
A_0 & = \frac{m_t^2}{M_W^2} \left[ |\vl|^2 + |\vr|^2 \right] \left(1 - x_W^2 \right)
+ \left[ |\gl|^2 + |\gr|^2 \right] \left(1 - x_W^2 \right) \notag \\
&  - 4 x_b \, \RE \left[ \vl \vr^* + \gl \gr^* \right]
- 2 \frac{m_t}{M_W} \RE \, \left[\vl \gr^* + \vr \gl^* \right]
\left(1 - x_W^2 \right)  \notag \\
& + 2 \frac{m_t}{M_W} x_b \,\RE \, \left[\vl \gl^* + \vr \gr^* \right]
\left(1 +x_W^2 \right) \,, \notag \\
B_0 & = \left[ |\vl|^2 + |\vr|^2 \right] \left(1 - x_W^2 \right) + \frac{m_t^2}{M_W^2} \left[ |\gl|^2 + |\gr|^2 \right] \left(1 - x_W^2 
 \right) \notag \\
& - 4 x_b \, \RE \left[ \vl \vr^* + \gl \gr^* \right]
- 2 \frac{m_t}{M_W} \RE \, \left[\vl \gr^* + \vr \gl^* \right]
\left(1 - x_W^2 \right) \notag \\
& + 2 \frac{m_t}{M_W} x_b \,\RE \, \left[\vl \gl^* + \vr \gr^* \right]
\left(1 +x_W^2 \right) \,,
\label{ec:AB}
\end{align}
$x_W = M_W/m_t$, $x_b = m_b/m_t$.
A naive combination of the current helicity fraction measurements~\cite{Abazov:2010jn,Aaltonen:2012tk,Aad:2012ky,CMS}, not taking into account correlations between different experiments and data sets, gives
\begin{eqnarray}
F_+ & = & 0.007 \pm 0.027 \,, \notag \\
F_0 & = & 0.659 \pm 0.042 \,,
\label{ec:Fexp}
\end{eqnarray}
in good agreement with the SM tree-level prediction $F_+ \simeq 3 \times 10^{-4}$, $F_0 = 0.697$, $F_- = 0.303$ for $m_t = 172.5$ GeV. This implies a relatively small asymmetry $A_\text{FB}^{tW} \simeq 0.11 P_z$. (Higher-order corrections~\cite{Czarnecki:2010gb} slightly modify these values, but the differences are well below the experimental uncertainty.)

\begin{figure}[htb]
\begin{center}
\epsfig{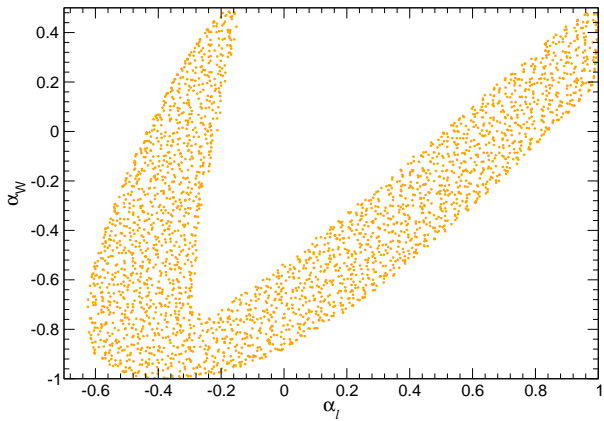}
\end{center}
\caption{Allowed region for $(\alpha_\ell,\alpha_W)$ at 68.3\% CL requiring agreement with current measurements of helicity fractions and LHC single top cross sections at 7 TeV.}
\label{fig:alab}
\end{figure}

When extracting the top polarisation from $A_\text{FB}^{tW}$ there will be, in addition to the experimental uncertainty associated to the measurement of this asymmetry, an uncertainty associated to the precision in the determination of helicity fractions. With the values in Eqs.~(\ref{ec:Fexp}), this uncertainty is $\Delta P_z / P_z = 0.12$, while with the foreseen LHC precision~\cite{AguilarSaavedra:2007rs} it will be reduced by a factor of two, $\Delta P_z / P_z = 0.06$. This can be compared to the ``theoretical'' uncertainty on $P_z$ that arises when extracting it from the distribution in Eq.~(\ref{ec:dist1}), which is associated to possible new physics in the decay.
For illustration, we show in Fig.~\ref{fig:alab} the estimated allowed region for $(\alpha_\ell,\alpha_W)$ at 68.3\% confidence level (CL). These limits are obtained with a combined fit to the four $Wtb$ couplings in Eq.~(\ref{ec:lagr}), using the code {\tt TopFit}~\cite{AguilarSaavedra:2006fy,AguilarSaavedra:2010nx} and taking as experimental data the helicity fractions in Eqs.~(\ref{ec:Fexp}), and the measured LHC $t$-channel single top cross section at 7 TeV, $\sigma = 68.5 \pm 5.8$ pb.\footnote{This value results from a weighted average of the ATLAS~\cite{Aad:2012ux} and CMS~\cite{:2012ep} results, ignoring correlations.
Tevatron measurements have much larger uncertainties, and LHC results at 8 TeV still have a worse precision.
In any case, even combining cross sections at different energies, it is hard to improve the limits beyond the 15\% level even with additional observables~\cite{AguilarSaavedra:2010nx}.}
Then, for $Wtb$ couplings within the range imposed by these measurements, the possible variation of the spin analysing powers is given by 
Fig.~\ref{fig:alab}. These limits are compatible with previous sensitivity projections \cite{AguilarSaavedra:2010nx} made under more optimistic assumptions and using additional observables not yet measured.
We observe that there is a large uncertainty on the spin analysing powers associated to possible new physics in the decay, that arises from (i) a number of parameters (four) in the effective vertex greater than the number of independent measurements (three); (ii) an approximate cancellation between $\vr$ and $\gl$ contributions to helicity fractions~\cite{AguilarSaavedra:2008gt}. Thus, even when the precision in the measurement of helicity fractions and single top cross sections is improved, this trend will persist. For example, using the charged lepton as spin analyser, the resulting uncertainty on the polarisation measurement is
\begin{equation}
P_z \in [-1,-1.6 P^\text{exp}_z] \cup [P_z^\text{exp},1] \,,
\end{equation}
being $P_z^\text{exp}$ the ``observed'' top quark polarisation (taken positive), extracted from Eq.~(\ref{ec:dist1}) assuming $\alpha_\ell = 1$. (The first of these intervals is empty if $P_z^\text{exp} > 0.62$.) Clearly, the larger the polarisation, the smaller will be this uncertainty. But, in any case, any measurement of $A_\text{FB}^{tW}$ will provide relevant information, and the best precision in top polarisation measurements will be achieved from combination of all observables available.

At the LHC, single top quarks produced in the $t$-channel process are highly polarised in the direction of the spectator quark~\cite{Mahlon:1999gz}, with $P_z \simeq 0.9$ for centre-of-mass energies $\sqrt s = 7,8$ TeV~\cite{AguilarSaavedra:2008gt}, so the expected asymmetry $A_\text{FB}^{tW} \simeq 0.1$ is measurable. LHC statistics are excellent and this measurement may eventually be dominated by systematics. In that case, for a precise determination of the single top polarisation it may be convenient to measure instead the ratio 
$A_\text{FB}^{tW} / A_\text{FB}$, being $A_\text{FB}$ the well-known lepton FB asymmetry in the $W$ rest frame, see Eq.~(\ref{ec:afb}). Given the present helicity fraction measurements, which imply $F_+ / F_- \lesssim 0.05$ (this ratio is $F_+ / F_- \simeq 10^{-3}$ in the SM), one has $A_\text{FB}^{tW} / A_\text{FB}\simeq -1/2\, P_z$ to a good approximation.  Alternatively, $A_\text{FB}^{tW}/ F_- \simeq 3/8\, P_z$ can also be measured. Besides, we note that the inequality (\ref{ec:ineq}) may also be used to obtain relevant bounds on $P_z$ in processes, such as single top production, where it is large. The product $\kappa_W \equiv \alpha_W P_z$ can be determined from the $\cos \theta_W$ distribution, see Eq.~(\ref{ec:dist1}). Then, if the experimental measurement is consistent with the SM prediction, say
$\kappa_W^\text{exp} \simeq 0.36$, the inequality  (\ref{ec:ineq}) implies\footnote{We are assuming $\alpha_W > 0$ here, in which case Eq.~(\ref{ec:ineq}) implies $1/\alpha_W \geq 1/(F_0-F_-+F_+)$. This assumption can explicitly be tested with the measurement of the sign of $P_z$ from $A_\text{FB}^{tW}$ and the sign of $\kappa_W^\text{exp}$.}
\begin{equation}
P_z \geq \frac{\kappa_W^\text{exp}}{F_0 - F_- + F_+} \simeq 0.9 \,,
\end{equation}
which is a very stringent bound since $P_z \leq 1$ by definition. Note, however, that the same result can be achieved with the measurement of $\kappa_\ell \equiv P_z \alpha_{\ell}$ from the $\cos \theta_\ell$ distribution, since $\alpha_\ell \leq 1$.

Finally, at a future $e^+ e^-$ International Linear Collider top quarks are produced in pairs with a small but non-zero polarisation, $P_z \simeq 0.14$ in the helicity axis for $\sqrt s = 500$ GeV~\cite{Grzadkowski:1996kn}. The precision expected for asymmetry measurements is excellent~\cite{Doublet:2012wf}, and therefore the measurement of this asymmetry may be very useful to complement top-spin-independent observables to probe anomalous contributions to $e^+ e^- \to t \bar t$ independently of the decay vertex~\cite{AguilarSaavedra:2012vh}.
In any case, it is clear that the new asymmetry $A_\text{FB}^{tW}$ introduced here provides a new handle to measure the top polarisation in any process, and test the presence of new physics in the top sector.

\section*{Acknowledgements}
This work has been supported by MICINN by projects FPA2006-05294 and FPA2010-17915, Junta de Andaluc\'{\i}a (FQM 101, FQM 03048 and FQM 6552) and Funda\c c\~ao
para a Ci\^encia e Tecnologia~(FCT) project CERN/FP/123619/2011.

\end{document}